# Giant magnetocaloric effect from reverse martensitic transformation in Ni-Mn-Ga-Cu ferromagnetic shape memory alloys

Sudip Kumar Sarkar[a,1], Sarita[a], P. D. Babu[b], Aniruddha Biswas[a], Vasudeva Siruguri[b] and Madangopal Krishnan[a]

[a]Glass and Advanced Materials Division, Bhabha Atomic Research Centre, Mumbai - 400085, India

[b]UGC-DAE Consortium for Scientific Research, Mumbai Centre, BARC, Mumbai - 400085, India

**Abstract:**

In an effort to produce Giant Magnetocaloric effect (GMCE) near room temperature, in a first ever such study, the austenite transformation temperature ($A_s$) was fine tuned to ferromagnetic Curie temperature ($T_C$) in Ferromagnetic Shape Memory Alloys (FSMA) and a large GMCE of $\Delta S_M$ = - 81.75 J/Kg-K was achieved in $Ni_{50}Mn_{18.5}Cu_{6.5}Ga_{25}$ alloy during reverse martensitic transformation (heating cycle) for a magnetic field change of 9 T at 302.5 K. Fine tuning of $A_s$ with $T_C$ was achieved by Cu substitution in $Ni_{50}Mn_{25-x}Cu_xGa_{25}$ ($0 \leq x \leq 7.0$)-based FSMAs. Characterizations of these alloys were carried out using Optical and Scanning Electron Microscopy, X-ray Diffraction (XRD), Differential Scanning Calorimetry (DSC) and DC magnetization measurements. Addition of Cu to stoichiometric Heusler type $Ni_2MnGa$ increases the martensitic transformation temperatures and decreases $T_C$. Concurrently, $\Delta S_M$ increases with Cu addition and peaks at 6.5 at. % Cu for which there is a virtual overlap between $T_C$ and $A_s$. Maximum Refrigerant Capacity (RCP) of 327.01 J/Kg was also achieved in the heating cycle for 9 T field change at 302.5 K. Corresponding values for the cooling cycle measurements (measured during forward transformation) were 30.4 J/Kg-K and 123.52 J/Kg respectively for the same 6.5 at. % Cu sample and same thermo-magnetic conditions.

[1]Corresponding author. Tel.: +91 22 2559 2390; fax: +91 22 2550 5151; e-mail: sudips@barc.gov.in

# 1. INTRODUCTION

Recent years have seen a significant rise in the research interest in magnetic refrigeration [1] at room temperature [2-4] to replace conventional gas compression-expansion based [5] refrigeration technology. Higher refrigeration efficiency, environment friendliness, low cost, small volume requirement, lack of toxicity and low noise production give magnetic refrigeration technology an edge over the others. As a result, there have been a large number of studies on various magnetic materials that show large Magneto Calorific Effect (MCE) near room temperature, which in turn can act as potential magnetic refrigerant material.

Giant MCE (GMCE) was first observed in $Gd_5(Si_{1-x}Ge_x)_4$ [2]. Following this, a number of different materials such as $La(Fe_xSi_{1-x})_{13}$ [6], $MnFeP_{1-x}As_x$ [7], $MnAs_{1-x}Sb_x$ [8] etc. were found to exhibit GMCE [3-6]. Ferromagnetic Shape Memory Alloys (FSMA), another class of functional materials, also show large MCE. Sharp change in magnetization associated with the structural transformation from high-temperature austenite (cubic) phase to low-temperature martensite phase of lower crystallographic symmetry, results not only in large MCE [9-11] in FSMAs, but also leads to many interesting functional properties like shape memory effect, magnetic field induced strain (MFIS) [11-15], magnetic field induced reverse transformation (MFIRT) [11] and kinetic arrest effect [16-18]. Coincidence of first order structural martensitic transformation and second order magnetic transition can turn MCE into GMCE. Now, there are two kinds of FSMAs, one that show conventional [9] MCE and the other that manifests inverse MCE [19-20]. Ni-Mn-Ga-based Heusler alloys are the most widely used alloys in this class of conventional magnetocaloric FSMAs. But, Ni-Mn-Ga-based Heusler alloys exhibit GMCE only in case of Ni-rich off-stoichiometric compositions [21]. Alternately, Cu substitution for Mn can also produce similar GMCE; for example, Stadler *et al* [9]˙ reported a maximum magnetic entropy change of $\Delta SM = -64$ J/Kg-K at 308 K for $Ni_2Mn_{0.75}Cu_{0.25}Ga$ polycrystalline alloys by tuning $M_s$ with $T_C$. Since then, many attempts were made to maximize the $\Delta SM$ value by exploring different possibilities in Ni-Mn-Cu-Ga system. Gautam *et al* [22] reported that small changes in Cu concentration in $Ni_2Mn_{0.75(1-x)}Cu_{0.25}Ga$ have virtually no effect in improving the magnetic entropy change. In the same line, Khan *et al* [23] reported that substitution of Fe and Ge in $Ni_2Mn_{0.75}Cu_{0.25}Ga$ also could not enhance MCE. There have been several other interesting studies on various different aspects of Ni-Mn-Ga-Cu quaternary system: shape memory effect, compressive plasticity and wide hysteresis [24], MFIRT [25], MFIS [26]. One of the most important studies on the effect of quaternary addition was performed by Kotaoka *et al* [27], who developed an extensive phase diagram by substituting Cu in Ni-Mn-Ga.

It may be noted here that in case of the Ni-Mn-Ga-based conventional MCE materials, the low temperature martensite phase has higher saturation magnetization as compared to the high temperature austenite phase [28]. The refrigeration cycle in these materials can be described as follows [29]. Application of magnetic field stabilizes the martensitic phase with higher saturation magnetization i.e, the material gets magnetized. The heat generation associated with this magnetization process is subsequently removed from the material by some circulating fluid. The material still remains in the magnetized martensitic state. In this condition, the material can extract heat from the surrounding if the magnetic field is withdrawn adiabatically and the material in turn, transforms back to the austenite phase. However, martensitic transformation is not completely thermo-elastic in nature, finite hysteresis is always present there [28, 30]. This makes the transformation path-specific. Therefore, for this first order magneto-structural transformation in conventional magnetocaloric FSMAs, the magnetic entropy change can be measured either when: (i) austenite transforms to martensite, during cooling; or alternately when (ii) martensite transforms to austenite, during heating. As a result, $T_C$ could be tuned either with $A_s$ or $M_s$. However, overlap of $T_C$ with $A_s$ is more efficient and promising than with $M_s$, since cooling in magnetic refrigeration is achieved during adiabatic withdrawal of magnetic field which is associated with the reverse transformation from martensite to austenite (reverse transformation starts at $A_s$); this corresponds to the heating cycle entropy change. Interestingly, nearly all previously reported magnetocaloric studies in FSMAs were focussed on tuning $T_C$ with $M_s$ so that GMCE could be achieved during the cooling cycle. There are just a couple of studies [31-32] in the literature on the heating cycle measurements. However, there too, no effort was made to get the $T_C$ to overlap with $A_s$ which could have enhanced the MCE during the heating cycle.

In view of this, the current study introduces a novel method of tuning $A_s$ with $T_C$. Our study explores the effect of Cu substitution in $Ni_{50}Mn_{25-x}Cu_xGa_{25}$ (x = 0, 1.25, 2.5, 3.75, 6.0, 6.2, 6.25, 6.5, 7.0) on the structural and magnetic transitions with an emphasis on tuning $T_C$ with $A_s$. The successful implementation of the idea would result in higher $\Delta SM$ in the heating cycle as compared to the cooling cycle.

## 2. EXPERIMENTAL

The buttons of the following alloys: $Ni_{50}Mn_{25-x}Cu_xGa_{25}$ (x = 0, 1.25, 2.5, 3.75, 6.0, 6.2, 6.25, 6.5, 7.0) were prepared by vacuum arc melting high purity (99.99 %) elements in appropriate proportions. Homogeneity was ascertained by re-melting the alloys multiple times. The buttons were sealed in a quartz ampoule filled with helium gas and solutionized at 1123 K for 24 h. Detailed characterization of these alloys was carried out using optical and Scanning Electron Microscopy (SEM), Energy Dispersive X-ray spectroscopy (EDX), X-Ray Diffraction (XRD), Differential Scanning Calorimetry (DSC) and DC magnetization techniques. Samples for metallography were etched using an aqueous solution of $FeCl_3$ in HCl. XRD experiments were carried out for both bulk and powder samples using a Cu Kα radiation. Isochronous DSC experiments were performed using a Mettler-Toledo machine at a rate of 10 K/min in argon atmosphere. DC magnetization was measured using a commercial 9 T PPMS-VSM (make Quantum Design).

## 3. RESULTS

### 3.1. Microstructure:

A single-phase microstructure was observed in as-solutionized condition for all the alloys. The microstructure shows austenite at room temperature for Cu-substituted $Ni_{50}Mn_{25-x}Cu_xGa_{25}$ (x = 0, 1.25, 2.5, 3.75, 6.0, 6.2, 6.25, 6.5, 7.0) alloys up to 6.5 at.% Cu. Above Cu:6.5, the microstructures are martensitic at room temperature. Fig. 1 shows the representative optical micrographs corresponding to (a) austenite and (b) martensite phase, respectively. The chemical compositions of these alloys were determined by EDX attached to an SEM. The compositions were found to be very close to the nominal compositions, as shown in Table 1.

### 3.2. Thermal analysis:

Fig. 2 shows the isochronal DSC plots of the ternary alloy $Ni_2MnGa$ and its quaternary counterparts, obtained by partial substitution of Mn by Cu, all in as-solutionized condition. All of them show clear evidence of reversible structural transformation which is martensitic in nature. The scan rate was 2 K/min for $Ni_2MnGa$ sample and 10 K/min for rest of the samples. The inset of Fig. 2 shows the Curie transitions in $Ni_2MnGa$ as well as in Cu:3.75 alloys, as evident in the DSC scans and are characterized by a change in the base line.

Characteristic martensitic transition (MT) temperatures are as follows: martensite start temperature ($M_s$), martensite finish temperature ($M_f$), austenite start temperature ($A_s$) and austenite finish temperature ($A_f$). These temperatures along with transformation enthalpies ($\Delta H$) for all the alloys are listed in Table 2. It is noticed that $M_s$ increases progressively with Cu addition while $T_C$ concomitantly decreases. Fig. 3 depicts the variation of $T_C$ and MT temperature as a function of Cu concentration. $T_C$, characterized by a change in the slope of the DSC plot, was determined for alloys up to Cu:3.75 where MT and $T_C$ were well apart. For all other alloys, separate signatures for MT and $T_C$ were not detected because of close proximity of these transitions and the weak heat evolution in magnetic transition. In addition, the enthalpy of transformation shows a gradual increase with increasing Cu concentration.

Similar trend was also noticed for the changes in entropy ($\Delta S$) associated with martensitic transformation. Entropy changes associated with the transformation ($\Delta S$) for the cooling as well as the heating processes were calculated by dividing the enthalpies of the forward and the reverse transformation respectively, by $T_0$, as shown in Table 2. $T_0$ is the transformation temperature where the parent phase has the same Gibbs energy as that of the product martensite phase and is approximated by $(A_f + M_s)/2$ [33-36].

*3.3. Phase analysis- XRD:*

Fig. 4(a) shows the typical X-Ray diffraction patterns of $Ni_{50}Mn_{25-x}Cu_xGa_{25}$ (x = 3.75, 6.5, 7.0) at room temperature. Addition of Cu to $Ni_2MnGa$ causes the microstructure to turn into martenstite (x > 6.5) from austenite (0 < x < 6.5) at room temperature. The austenite phase has ordered cubic structure, whereas the martensite is non-modulated type having tetragonal crystal structure [22, 37-38]. Fig. 4(b) represents the diffraction pattern of Cu:7.0 sample which has martensitic microstructure at room temperature. Structure reveals mostly NM type martensite, in addition to that a trace amount of parent phase and martensite of type 7M was also observed. XRD results are in agreement with our optical microscopy observations.

### 3.4. Magnetization measurement:

Fig.s 5(a), (b) and (c) show representative low field (0.01 T) magnetization versus temperature data recorded for the different alloy compositions in zero-field cooled (ZFC), field-cooled cooling (FCC) and field-cooled warming (FCW) sequences. All these alloys show similar thermo-magnetic behaviors. Starting with high temperature austenite phase, with decreasing temperature, a sharp increase of magnetization was observed corresponding to the paramagnetic to ferromagnetic transition, $T_C$ of austenite phase. Further decreasing of temperature causes a sudden drop in magnetization, which corresponds to martensitic transformation. Thus, the transformation sequence is: paramagnetic austenite → ferromagnetic austenite → ferromagnetic martensite. $T_C$ has been determined by identifying the temperature at which ZFC magnetization rises sharply when cooled from higher temperature (i.e., the point at which the temperature derivative of ZFC curve in this region shows a maximum). On the other hand, the reverse martensitic transformation has been confirmed by the sudden jump in magnetization upon heating. Between the cooling and the heating processes, there is an obvious thermal hysteresis around the MT temperatures, which is a signature of the first-order transition. Ternary $Ni_{50}Mn_{25}Ga_{25}$ alloy shows one pre-martensitic transformation [8] at 240 K which is not present in any other alloys in the Ni-Mn-Ga-Cu quaternary series. Addition of Cu brings MT and $T_C$ closer as can be seen clearly in M vs. T plots. Above 6 at.% of Cu, M vs. T plot shows step like appearance around MT, typical characteristic of first order magneto-structural transition. Fig. 5(c) is the representative plot for those step like magneto-structural transitions shown for Cu:6.5. In this case, paramagnetic austenite directly transforms to ferromagnetic martensite upon cooling. The transformation sequence here is: paramagnetic austenite → ferromagnetic martensite. As a result, the change in magnetization ($\Delta M$) is higher here in comparison to the previously discussed one, which reflects directly in MCE.

Fig. 6(a) shows typical M vs. H curves for Cu:6.5 alloy measured during the cooling cycle. The curve corresponding to 305 K shows magnetic field induced austenite to martensite transformation at about 6.8 T field. At 300 K, the same phenomenon was observed for lower critical field of just 1.8 T. Similar behaviour was also seen for the rest of the alloys. M vs. H plot for Cu:3.75 alloy, shown in Fig. 6(b), exhibits a slightly strange behaviour. Magnetization of the parent phase is more easily saturated than that of the martensitic phase, because the martensitic phase has higher magneto-crystalline anisotropy energy than the parent phase [39– 40]. In the low magnetic field region, magnetization of the parent phase is higher than the martensitic phase. The magnetizations become equal at approximately 0.2 – 0.5 T, and subsequently at higher magnetic fields, the magnetization of the martensite phase becomes higher than that of the parent austenite phase. Saturation magnetization was seen to decrease with Cu addition.

Isothermal M vs. H plots were used for MCE calculation. MCE is measured either in terms of the isothermal magnetic entropy change ($\Delta SM$) or adiabatic temperature change ($\Delta T_{ad}$). Detailed thermodynamics of the MCE can be found in the work by Pecharsky *et al* [2]. A number of M vs. H isotherms at close interval of 5 K (at 2.0 K for temperature close to $T_C$ and $M_s$) were recorded on either side of $T_C$ to calculate the MCE for all the compositions. The changes in entropy ($\Delta SM$) for both the cooling and heating cycles are evaluated as a measure of the MCE from these isotherms using the following relationship [36]:

$$\Delta SM(T)_{\Delta H} = \int_{H_I}^{H_F} \left(\frac{\partial M(T,H)}{\partial T}\right)_H dH \qquad (1)$$

where, M is magnetization, H is magnetic-field strength and T is absolute temperature [2]. Fig. 7 shows the typical evolution of the entropy change ($\Delta SM$) for different applied fields over a wide temperature range encompassing $T_C$ for one of the alloys (Cu:6.5) for both the thermodynamic cycles. $\Delta SM$ is negative for heating cycle because dM/dT is negative as austenite has lower saturation magnetization values. For the sake of clarity, the Fig. 7 shows the results for only some selected field values. Fig. 8 shows the comparative plots of evolution of $\Delta SM$ with temperature for all the compositions for magnetic field change of 9 T for heating cycle. The inset of Fig. 8 presents the same for cooling cycle. It clearly shows that $\Delta SM$ increases with increasing Cu substitution and Cu:6.5 alloy records the maximum value. The maximum changes in entropy for all the compositions are listed in Table 3.

Refrigeration capacity (RCP), the amount of heat transferred in one thermodynamic cycle, is another important parameter to evaluate magnetic refrigeration. RCP was calculated by integrating the $\Delta SM$ (T,H)$_H$ curves over the full-width at half maximum (FWHM) using the relation [41-43]:

$$RCP = \int_{T_1}^{T_2} \Delta SM(T,H)_H \, dT \qquad (2)$$

Fig. 9 shows the variation of RCP and FWHM with Cu concentration (at.%) for all the alloys. Cu:6.5 was seen to have the highest refrigeration capacity of 327.01 J/Kg at 302.5 K. FWHM was found to remain almost constant, valued around 4 K, for all the alloys.

## 4. DISCUSSIONS

Substitution of Cu for Mn in ternary Ni$_2$MnGa causes MT temperature to increase and T$_C$ to decrease as is shown in Fig. 3. MT temperature increases because Cu substitution results in an enhancement of Ni covalency, and therefore, leads to a stronger Ni-Ga chemical bond as indicated by the increase in valence electron concentration per atom (e/a) with Cu addition. Since the martensitic transition is related to the formation of Ni *d* and Ga *p* hybrid states [44-45], a higher degree of Ni-Ga hybridization in Ni$_{50}$Mn$_{25-x}$Cu$_x$Ga$_{25}$ ($0 \leq x \leq 7.0$) makes the chemical bond stronger; as a result, more energy is required to trigger the martensitic transition. On the other hand, T$_C$ decreases with addition of Cu which is because of the following two possible reasons. First reason is the decrease in magnetic atom i.e., Mn concentration with addition of Cu. Secondly, addition of Cu can change the inter-atomic distance between Mn atoms leading to weakening of ferromagnetic exchange interaction. In addition, Cu substitution also changes the microstructure at room temperature from austenite to martensite, for alloys containing more than 6.5 at.% Cu. The main challenge of the current work was to arrive at an appropriate alloy composition where A$_s$ and T$_C$ would be as close as possible. In general, M$_s$ and A$_s$ differ by over a couple of degrees around the martensitic transition due to thermal hysteresis since martesitic transformation is not completely a thermo-elastic one in Ni-Mn-Ga system [46]. Both the phases, austenite and martensite coexist in that hysteresis region. The equilibrium between the austenite and the martensite at an interface is defined by the following equation [47]:

$$\Delta G_{chem}^{M-A} - \Delta G_{elast}^{M-A} + E_{irr}^{M-A} = 0 \qquad (3)$$

where, $\Delta G_{chem}^{M-A}$ is the chemical free energy, $\Delta G_{elast}^{M-A}$ is the elastic strain energy that arises from lattice mismatch and volume fraction difference between martensite and austenite. The transformation process is associated with an irreversible energy loss, $E_{irr}^{M-A}$, which is the origin of hysteresis effect at martensitic transition.

Our result clearly shows that though all the $Ni_{50}Mn_{25-x}Cu_xGa_{25}$ ($6.0 \leq x \leq 7.0$) samples show first order magneto-structural transformation, only the virtual overlapping of $A_s$ with $T_C$ for Cu:6.5 sample results in the highest magnetic entropy change of - 81.75 J/Kg-K at 302.5 K by magnetic field change of 9 T during the heating cycle. In contrast, cooling cycle results in a magnetic entropy change of only 30.4 J/Kg-K for the same alloy with the same thermo-magnetic conditions. Fig. 10 provides a graphical insight of the dependence of magnetic entropy change for heating cycle with ($T_C - A_s$) in $Ni_{50}Mn_{25-x}Cu_xGa_{25}$ alloys. A strong correlation between the difference ($T_C - A_s$) and $\Delta SM$ was found. This correlation means that a larger $\Delta SM$ can be obtained from a specimen with a smaller ($T_C - A_s$) for heating cycle measurements. Inset of Fig. 10 shows a similar kind of dependency for magnetic entropy with ($T_C - M_s$) for the cooling cycle. Similar strong correlation between the $\Delta SM$ and ($T_C - M_s$) has been reported earlier in a few other FSMAs, such as the Ni-Mn-In and Ni-Co-Mn-In alloys for cooling cycle measurements [35]. Naturally, in the current case, magnetic entropy change increases systematically for heating cycle with Cu addition, since the difference between $A_s$ and $T_C$ declines progressively. A comparison of MCE across several well-known materials is shown in Table 4. The entropy change recorded by Cu:6.5 (60.22 J/Kg-K for 5 T field change) alloy during heating cycle ($A_s \sim T_C$) is very close to the value of - 65 J/Kg-K which is the highest ever reported value in polycrystalline material measured in a cooling cycle ($M_s \sim T_C$) for 5 T field change in $Ni_2Mn_{0.75}Cu_{0.25}Ga$, as shown by Stadler *et al* [9]. However, the RCP in our case shows a significant increase over the previously reported value in Ni-Mn-Ga FSMA series. The value of RCP enhanced to 94.56 J/Kg for Cu:6.5 alloy as compared to 84 J/Kg in case of $Ni_2Mn_{0.75}Cu_{0.25}Ga$ [9] for the same magnetic field change of 5 T by virtue of higher FWHM.

## 5. CONCLUSIONS

The primary findings of the current study can be summarized as follows:

1) Cu substitution in ternary Ni$_2$MnGa causes MT temperature to increase as valence electron concentration per atom (e/a) increases; and it causes T$_C$ to decrease because Mn is the atom which carries the magnetic moment in this type of Heusler alloys which Cu replaces.

2) Cu addition also changes the microstructure from austenite to martensite at room temperature above 6.5 at.% Cu concentration.

3) A virtual overlapping of A$_s$ with T$_C$, found in Ni$_{50}$Mn$_{18.5}$Cu$_{6.5}$Ga$_{25}$, gives rise to the highest magnetic entropy change of - 81.75 J/Kg-K at 302.5 K by magnetic field change of 9 T during the heating cycle. In contrary, cooling cycle results in only 30.4 J/Kg-K for the same alloy with the same thermo-magnetic condition.

4) Highest magnetic refrigerant capacity has been recorded for Cu: 6.5 sample of 94.56 J/Kg for magnetic field change of 5 T among Ni-Mn-Ga based FSMA series.

5) Higher ΔSM and refrigeration capacity from heating cycle measurements compared to cooling cycle by tuning T$_C$ with A$_s$ is very promising and can be successfully applied for magnetic refrigeration process for conventional magnetocaloric FSMAs.

## REFERENCES

[1] C Zimm, C Jastrab, A Sternberg, V K Pecharsky, K A Gschneidner, Jr., M Osborne, and I Anderson, Adv. Cryog. Eng. 43 (1998) 1759

[2] V K Pecharsky, and K A Gschneidner, Jr., Phys. Rev. Lett. 78 (1997) 4494

[3] V Provenzano, A J Shapiro and R D Shull, Nature (London) 429 (2004) 853

[4] O Tegus, E Brück, K H J Buschow and R D de Boer, Nature (London) 415 (2002) 150

[5] J Glanz, Science 279 (1998) 2045

[6]  G H Wen, R K Zheng, X X Zhang, W H Wang, J L Chen and G H Wu, J. Appl. Phys. 91 (2002) 8537

[7]  E Brück, M Ilyn, A M Tishin and O Tegus, J. Magn. Magn. Mater. 290 (2005) 8

[8]  H Wada, T Morikawa, T Taniguchi, T Shibata, Y Yamada and Y Akishige, Physica B 328 (2003) 114

[9]  S Stadler, M Khan, J Mitchell, N Ali, A M Gomes, I Dubenko, A Y Takeuchi and A P Guimarães, Appl. Phys. Lett. 88 (2006) 192511

[10] V K Pecharsky, K A Gschneidner, Jr., A O Pecharsky and A M Tishin, Phys. Rev. B 64 (2001) 144406

[11] K Ullakko, J K Huang, C Kantner, R C O'Handley and V V Kokorin, Appl. Phys. Lett. 69 (1996) 1966

[12] R Kainuma , Y Imano, W Ito, Y Sutou, H Morito, S Okamoto, O Kitakami, K Oikawa, A Fujita, T Kanomata and K Ishida, Nature 439 (2006) 957

[13] R D James and M Wuttig, Phil. Mag. A 77 (1998) 1273

[14] H X Zheng, M X Xia, J Liu, Y L Huang and J G Li, Acta Mater. 53 (2005) 5125

[15] S Kustov, M L Corro´, E Cesari, J I Pe´rez-Landaza´bal and V Recarte, J. Phys. D: Appl. Phys. 43 (2010) 175002

[16] W Ito, I Kouhei, R Y Umetsu, R Kainuma, K Koyama, K Watanabe, A Fujita, K Oikawa, K Ishida and T Kanomata, Appl. Phys. Lett. 92 (2008) 021908

[17] X Xu, W Ito, M Tokunaga, R Y Umetsu, R Kainuma and K Ishida, Mater. Trans. 51 (2010) 1357

[18] V Siruguri, P D Babu, S D Kaushik, A Biswas, S K Sarkar and M Krishnan, J. Phys.: Condens. Matter 25 (2013) 496011

[19] Y Sutou, Y Imano, N Koeda, T Omori, R Kainuma, K Ishida and K Oikawa, Appl. Phys. Lett. 85 (2004) 4358

[20] R Kainuma, Y Imano, W Ito, H Morito, Y Sutou, K Oikawa, A Fujita, K Ishida, S Okamoto, O Kitakami and T Kanomata, Appl. Phys. Lett. 88 (2006) 192513

[21] M Pasquale, Phys. Rev. B 72 (2005) 0944351

[22] B R Gautam, I Dubenko, J C Mabon, S Stadler and N Ali, J. Alloys Compd. 472 (2009) 35

[23] M Khan, S Stadler and N Ali, J. Appl. Phys. 101 (2007) 09C515

[24] J Wang and C Jiang, Scr. Mater. 62 (2010) 298

[25] P Li, J Wang, C Jiang and H Xu, J. Phys. D: Appl. Phys. 44 (2011) 285002

[26] C Jiang, J Wang, P Li, A Jia and X Xu, Appl. Phys. Lett. 95 (2009) 012501

[27] M Kataoka, K Endo, N Kudo, T Kanomata, H Nishihara, T Shishido, Y R Umetsu, M Nagasako and R Kainuma, Phys. Rev. B 82 (2010) 214423

[28] V Basso, C P Sasso, K P Skokov, O Gutfleisch and V Khovaylo, Phys. Rev. B 85 (2012) 014430

[29] V K Pecharsky and K A Gschneidner, Jr., J. Magn. Magn. Mater. 200 (1999) 44

[30] T Krenke, S Aksoy, E Duman, M Acet, X Moya, L Manosa and A Planes, J. Appl. Phys. 108 (2010) 043914

[31] J F Duan, P Huang, H Zhang, Y Long, G H Wu, R C Ye, Y Q Chang and F R Wan, J. Alloys Compd. 441 (2007) 29

[32] B Bo, L Yi, J Duan, W Guangheng, Y Rongehng and W Farong, J. Rare Earths 27 (2009) 875

[33] V V Khovailo, K Oikawa, T Abe and T Takagi, J. Appl. Phys. 93 (2003) 8483

[34] W Ito, Y Imano, R Kainuma, Y Sutou, K Oikawa and K Ishida, Metall. Mater. Trans. 38A (2007) 759

[35] E Cesari, D Salas and S Kustov, Mater. Sci. Forum. 684 (2011) 49

[36] S K Sarkar, A Biswas, P D Babu, S D Kaushik, A Srivastava, V Siruguri and M Krishnan, J. Alloys Compd. 586 (2014) 515

[37] Y Ma, S Awaji, K Watanabe, M Matsumoto and N Kobayashi, Solid State Commn. 113 (2000) 671

[38] J Pons, V A Chernenko, R Satamarta and E Cesari, Acta Mater. 48 (2000) 3027

[39] A A Cherechukin, T Takagi, M Matsumoto and V D Buchel'nikov, Phys. Lett. A 326 (2004) 146

[40] P J Brown, J Crangle, T Kanomata, M Matsumoto, K U Neumann, B Ouladdiaf and K R A Ziebeck, J. Phys.: Condens. Matter 14 (2002) 10159

[41] R Tickle and R D James, J. Magn. Magn. Mater. 195 (1999) 627

[42] J Chen, Z Hana, B Qiana, P Zhang, D Wang and Y Duc, J. Mag. Mag. Mater. 323 (2011) 248

[43] V K Sharma, M K Chattopadhyay, R Kumar, T Ganguli, P Tiwari and S B Roy, J. Phys.: Condens. Matter 19 (2007) 496207

[44] A Rostamnejadi, M Venkatesan, P Kameli, H Salamati and J Coey, J. Magn. Magn. Mater. 323 (2011) 2214

[45] A T Zayak, P Entel, K M Rabe, W A Adeagbo and M Acet, Phys. Rev. B 72 (2005) 054113

[46] P J Brown, A P Gandy, T Kanomata, M Matsumoto, K Neumann, K U Neumann, A Sheikh and K R A Ziebeck, Mater. Sci. Forum. 583 (2008) 285

[47] P J Shamberger and F S Ohuchi, Phys. Rev. B 79 (2009) 144407

**Highlights:**

- In a first ever such study, the austenite transformation temperature ($A_s$) was fine tuned to ferromagnetic Curie temperature ($T_C$) in Ferromagnetic Shape Memory Alloys (FSMA) to produce Giant Magnetocaloric effect (GMCE) near room temperature from reverse martensitic transformation.

- Fine tuning of $A_s$ with $T_C$ was carried out by Cu substitution in $Ni_{50}Mn_{25-x}Cu_xGa_{25}$ ($0 \leq x \leq 7.0$)-based FSMAs.

- Cu addition also changes the microstructure from austenite to martensite at room temperature above 6.5 at. % Cu concentration.

- Martensitic transformation temperature increases with Cu addition while $T_C$ decreases.

- A virtual overlapping of $A_s$ with $T_C$, found in $Ni_{50}Mn_{18.5}Cu_{6.5}Ga_{25}$, gives rise to the highest magnetic entropy change of - 81.75 J/Kg-K at 302.5 K by magnetic field change of 9 T during the heating cycle (reverse martensitic transformation). In contrary, cooling cycle (forward transformation) results in only 30.4 J/Kg-K for the same alloy with the same thermo-magnetic condition.

- Highest magnetic refrigerant capacity has been recorded for Cu: 6.5 sample of 94.56 J/Kg for magnetic field change of 5 T among Ni-Mn-Ga based FSMA series.

**Table**

# List of Tables

**Table 1**. EDX results showing chemical composition of $Ni_{50}Mn_{25-x}Cu_xGa_{25}$ alloys.

| (at.%) | e/a | Ni | Mn | Cu | Ga |
|---|---|---|---|---|---|
| Cu:0 | 7.59 | 51.12 | 25.19 | ------- | 23.69 |
| Cu:3.75 | 7.77 | 51.69 | 21.24 | 3.72 | 23.35 |
| Cu:6.0 | 7.75 | 49.67 | 19.6 | 6.07 | 24.66 |
| Cu:6.2 | 7.76 | 49.62 | 19.67 | 6.23 | 24.48 |
| Cu:6.25 | 7.74 | 49.39 | 19.44 | 6.28 | 24.89 |
| Cu:6.5 | 7.77 | 49.89 | 19.1 | 6.43 | 24.58 |
| Cu:7.0 | 7.82 | 50.1 | 18.6 | 7.1 | 24.1 |

**Table 2.** Martensitic transition temperatures of the alloys obtained from DSC scans.

| Alloy | $M_s$ K | $M_f$ K | $A_s$ K | $A_f$ K | $T_C$ K | $\Delta H^{M \to A}$ J/gm | $\Delta H^{A \to M}$ J/gm | $\Delta T$ ½($M_s$+ $A_f$) K | $\Delta SH$ J/Kg-K | $\Delta SC$ J/Kg-K |
|---|---|---|---|---|---|---|---|---|---|---|
| Cu:0 | 182.3 | 173.8 | 187.2 | 194.7 | 364.3 | -1.05 | 0.7 | 188.5 | 5.57 | 3.71 |
| Cu:3.75 | 256.4 | 248.4 | 264.3 | 271.8 | 323.16 | -3.96 | 4.1 | 264.1 | 14.99 | 15.52 |
| Cu:6.0 | 279.5 | 272.3 | 284.7 | 292.5 | 295.06 | -4.23 | 4.18 | 286.0 | 14.79 | 14.61 |
| Cu:6.2 | 283.4 | 265.6 | 281.3 | 302.3 | 294.4 | -6.0 | 5.83 | 292.9 | 20.48 | 19.91 |
| Cu:6.25 | 287.8 | 275.4 | 286.4 | 299.4 | 295.2 | -6.65 | 5.55 | 293.6 | 22.65 | 18.90 |
| Cu:6.5 | 288.5 | 281.6 | 291.0 | 299.4 | 292.9 | -7.23 | 5.94 | 294.0 | 24.77 | 20.21 |
| Cu:7.0 | 299.6 | 288.5 | 304.7 | 318.1 | 312.0 | -8.50 | 7.7 | 308.9 | 27.52 | 24.93 |

**Table 3.** Magnetic entropy (Δ*SM*) changes for all the alloys with fields.

| Sample | Temperature in K where Δ*SM* is maximum | Δ*SM* (J/Kg-K) with change in field up to | | | | | |
|---|---|---|---|---|---|---|---|
| | | 9T | 5T | 1T | 9T | 5T | 1T |
| | | Heating | | | Cooling | | |
| Cu:3.75 | 269.5 | 7.6 | 3.98 | 0.43 | 5.68 | 1.65 | 0.23 |
| Cu:6.0 | 292.0 | 28.53 | 24.42 | 4.86 | 17.20 | 15.36 | 0.92 |
| Cu:6.2 | 294.0 | 31.17 | 22.06 | 5.15 | ------- | ------- | ------- |
| Cu:6.25 | 299.5 | 43.88 | 33.51 | 10.2 | ------- | ------- | ------- |
| Cu:6.5 | 302.5 | 81.75 | 60.22 | 10.16 | 30.40 | 24.30 | 11.10 |
| Cu:7.0 | 316.25 | 46.34 | 31.89 | 6.23 | 45.0 | 28.94 | 4.17 |

**Table 4.** Comparison of magnetic entropy change between various MCE materials.

| Material | $-\Delta S_M$ (J/kg-K) | Tmax (K) | $\Delta H$ (Tesla) | Notes | MCE |
|---|---|---|---|---|---|
| Gd metal | <5 | 290 | 5 | Classic MCE material | Cooling cycle |
| $Gd_5(Si_{1-x}Ge_x)_4$ | 18.6 | 300 | 5 | Ref. [1] | |
| MnFe $P_{0.45}As_{0.55}$ | 18 | 310 | 5 | Ref. [2] | |
| $Ni_2Mn_{0.75}Cu_{0.25}Ga$ | 64 | 308 | 5 | Ref. [9] | |
| $Ni_{2+x}Mn_{1-x}Ga$ (Polycrystalline) | 68 | 350 | 5 | Ref. [4] | |
| $Ni_2Mn_{0.74}Cu_{0.26}Ga$ (Polycrystalline) | 60.22 | 302.5 | 5 | Present work | Heating Cycle |
| | 81.75 | 302.5 | 9 | | |

**Figure**

**Figures**

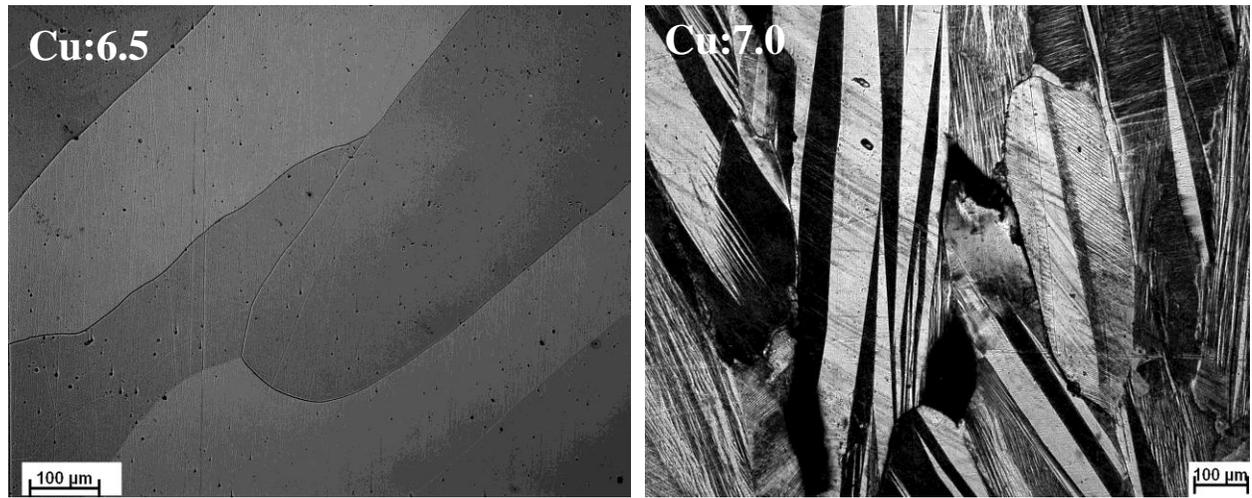

**Fig. 1.** Evolution of room temperature martensitic microstructure from parent with partial substitution of Mn by Cu in $Ni_{50}Mn_{25-x}Cu_xGa_{25}$ alloy: (a) Mn:6.5; (b) Cu:7.0.

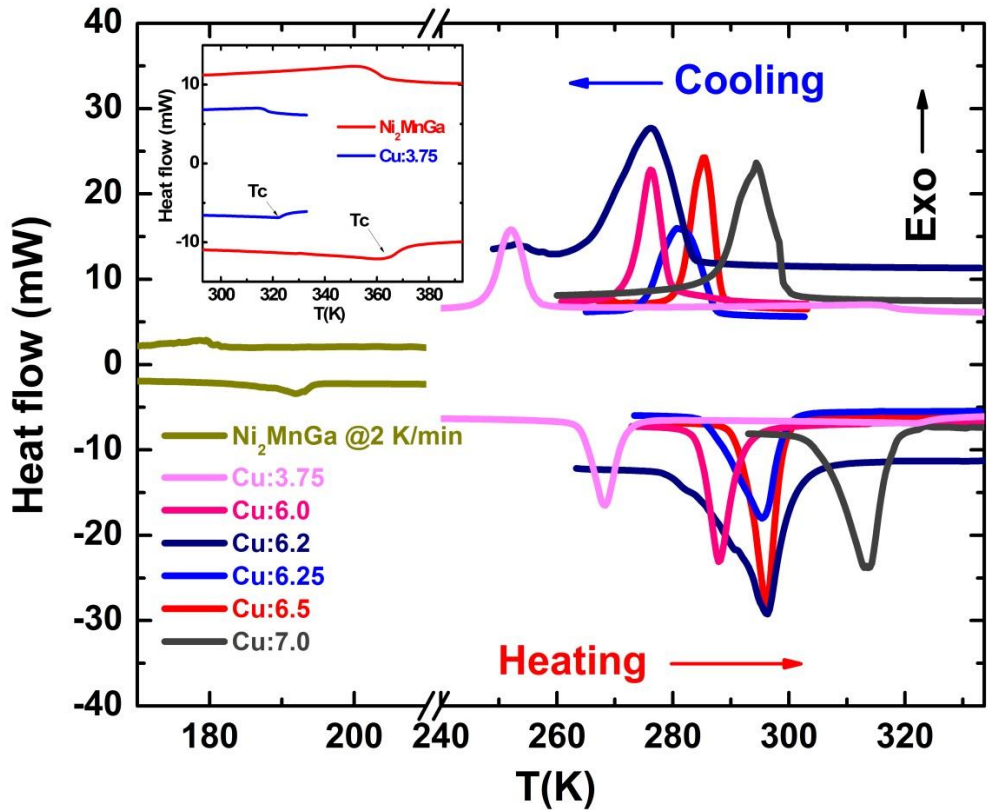

**Fig. 2.** Effect of Cu addition on martensitic transformation as seen in DSC scans and inset shows Curie transitions.

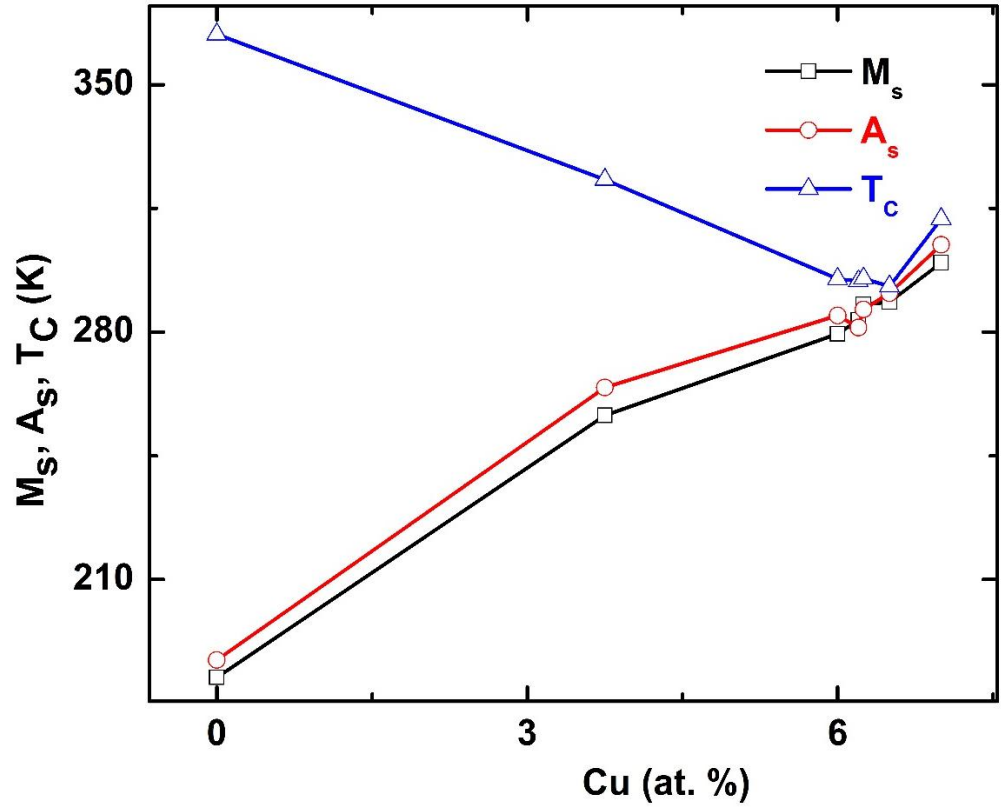

**Fig. 3.** Variation of $T_C$ and MT temperature as function of Cu concentration.

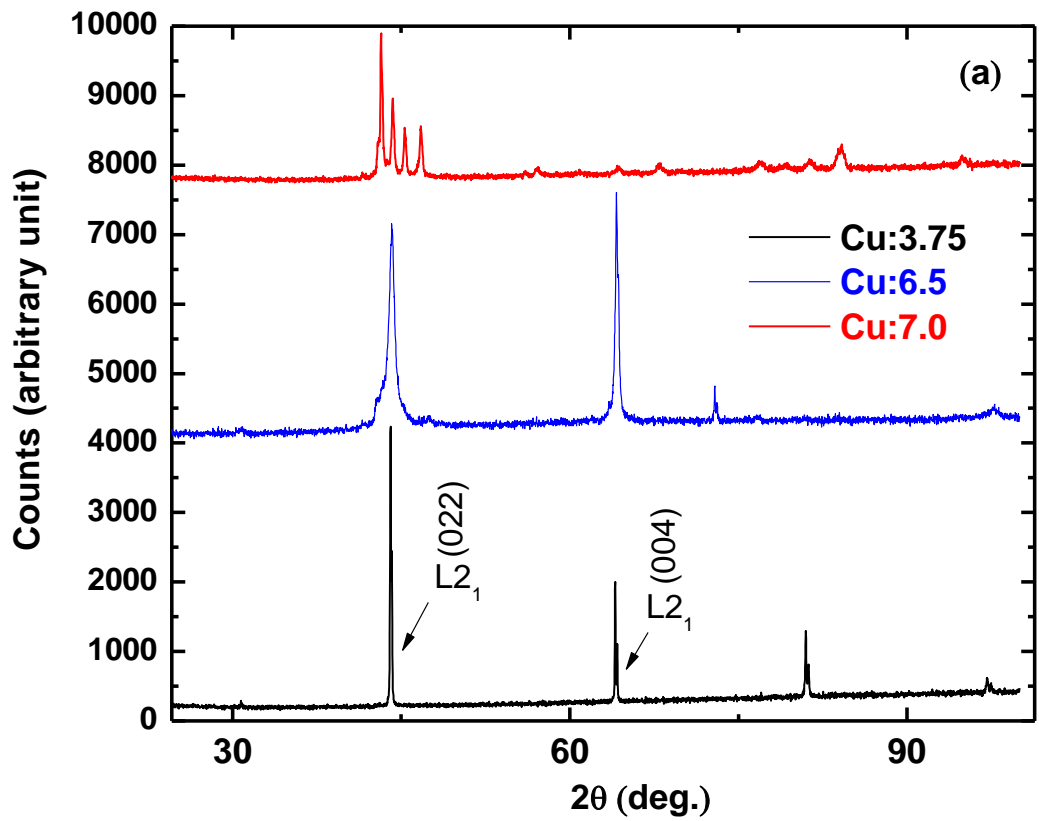

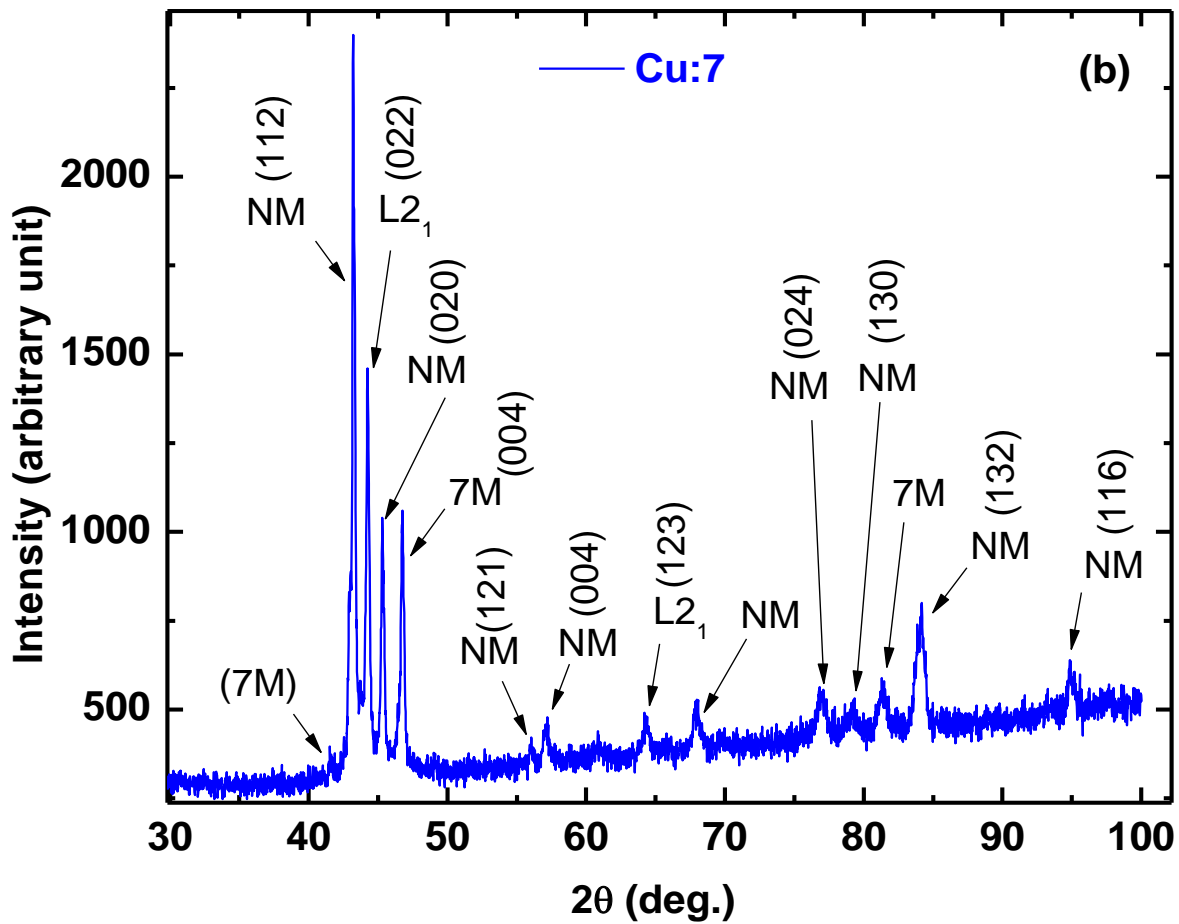

**Fig. 4.** XRD pattern for: (a) evolution of martensitic phase from $L2_1$ phase with increasing Cu addition; (b) mostly tetragonal NM type martensite observed in Cu:7.0 sample, taken as example.

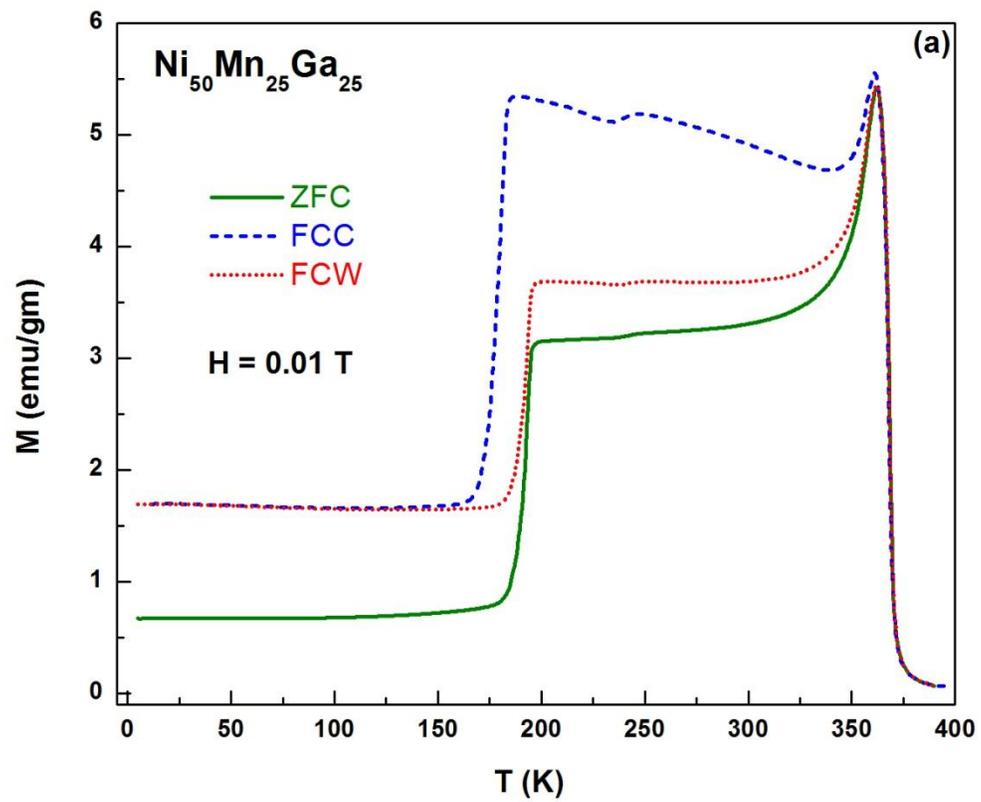

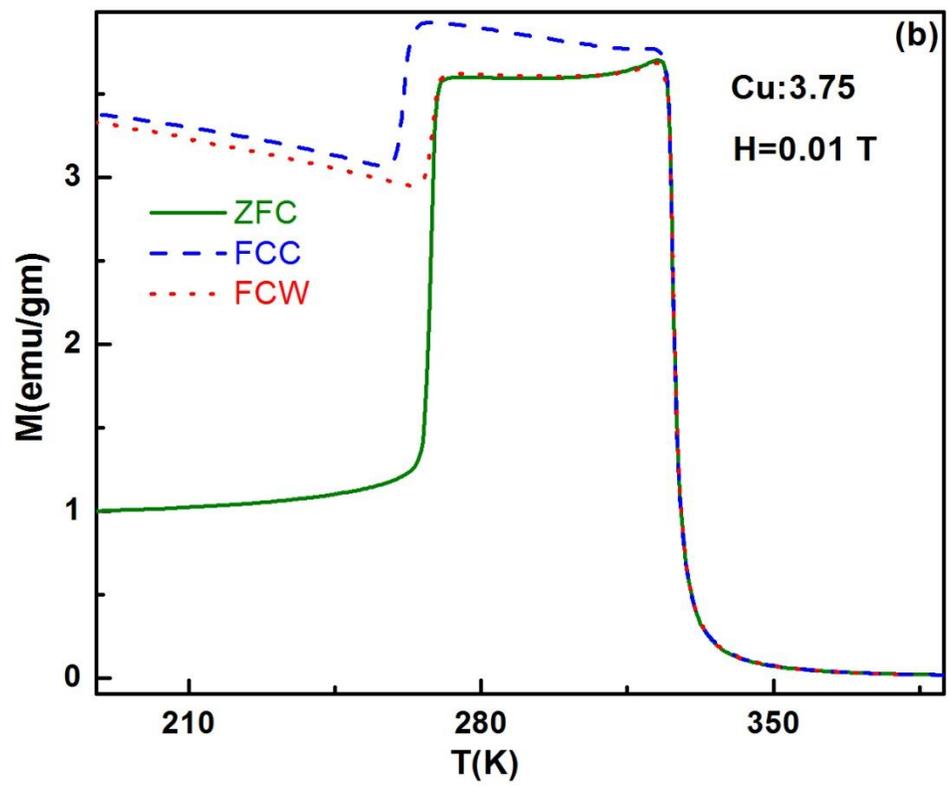

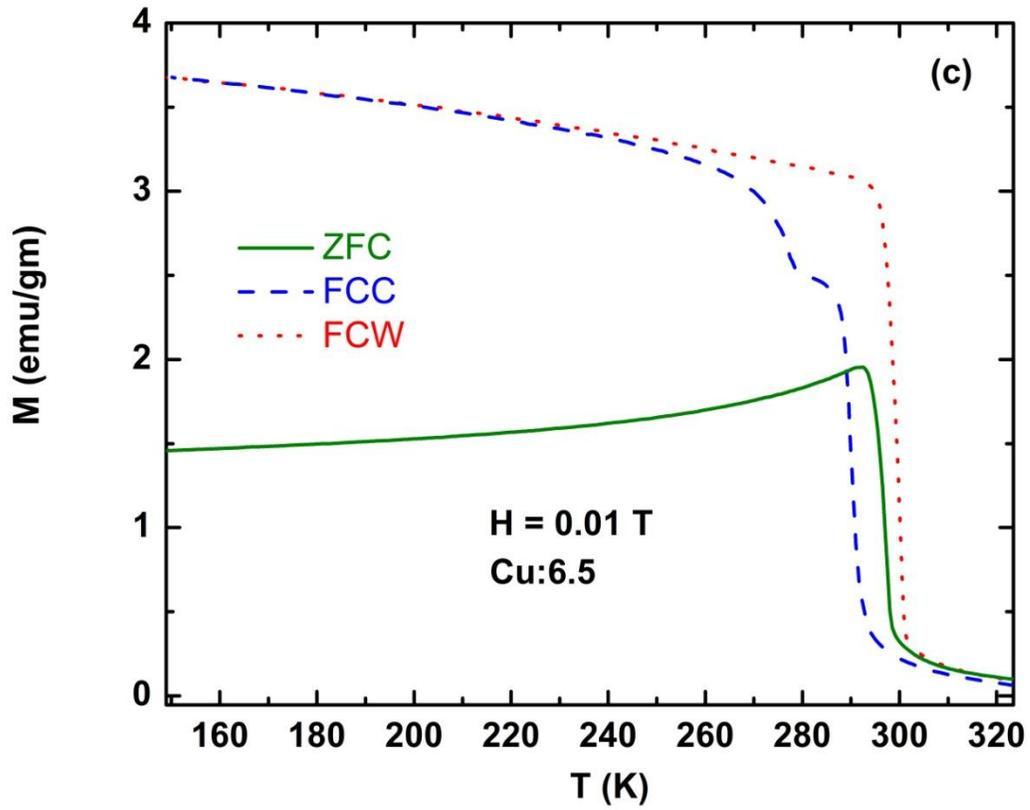

**Fig. 5.** Representative Zero Field Cooled (ZFC), Field-Cooled Cooling (FCC) and Warming (FCW) magnetization curves for: (a) $Ni_{50}Mn_{25}Ga_{25}$ (b) Cu:3.75 (c) Cu:6.5 at H = 0.01 T.

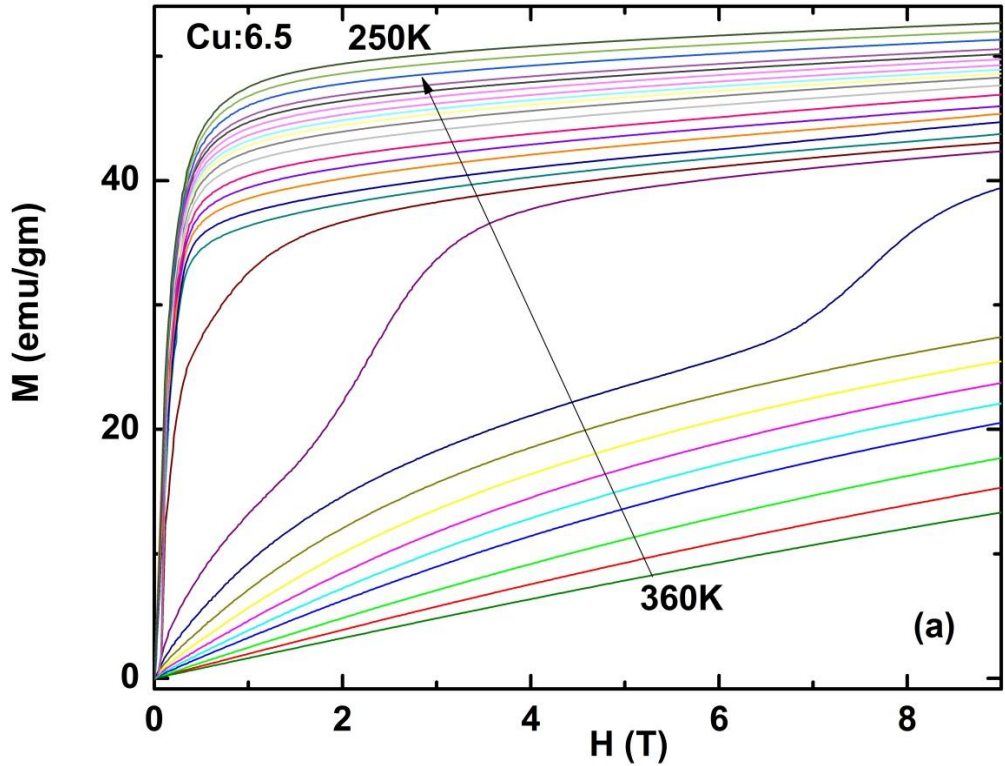

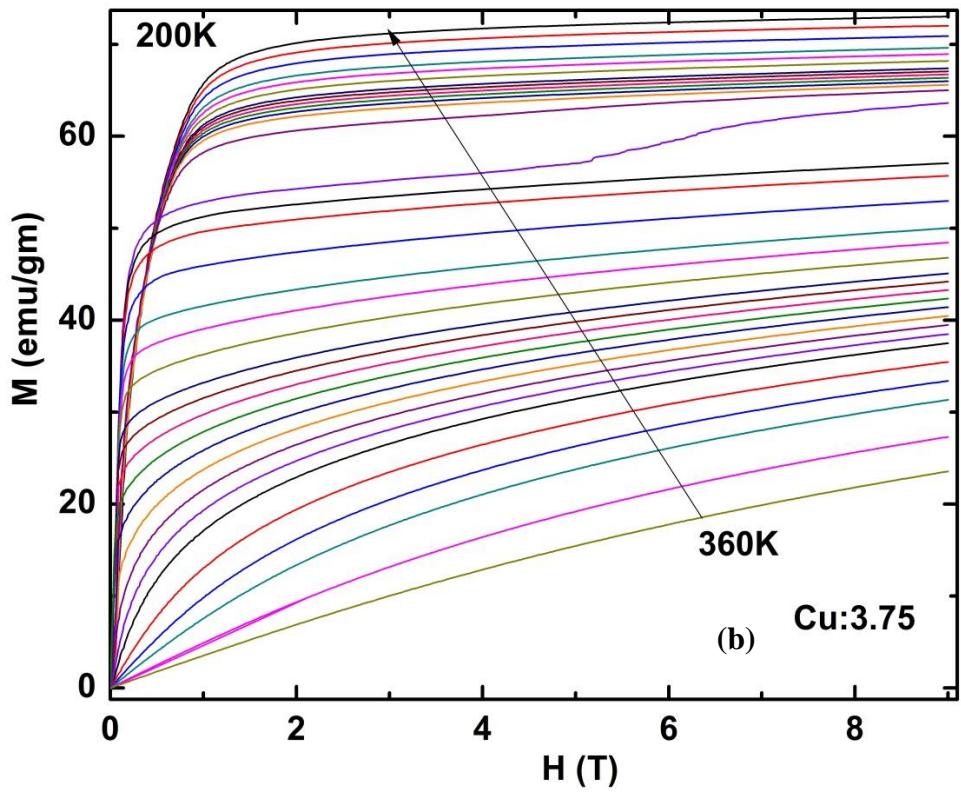

**Fig. 6.** Representative M vs. H isotherms shown for (a) Cu:6.5 and (b) Cu:3.75 alloy.

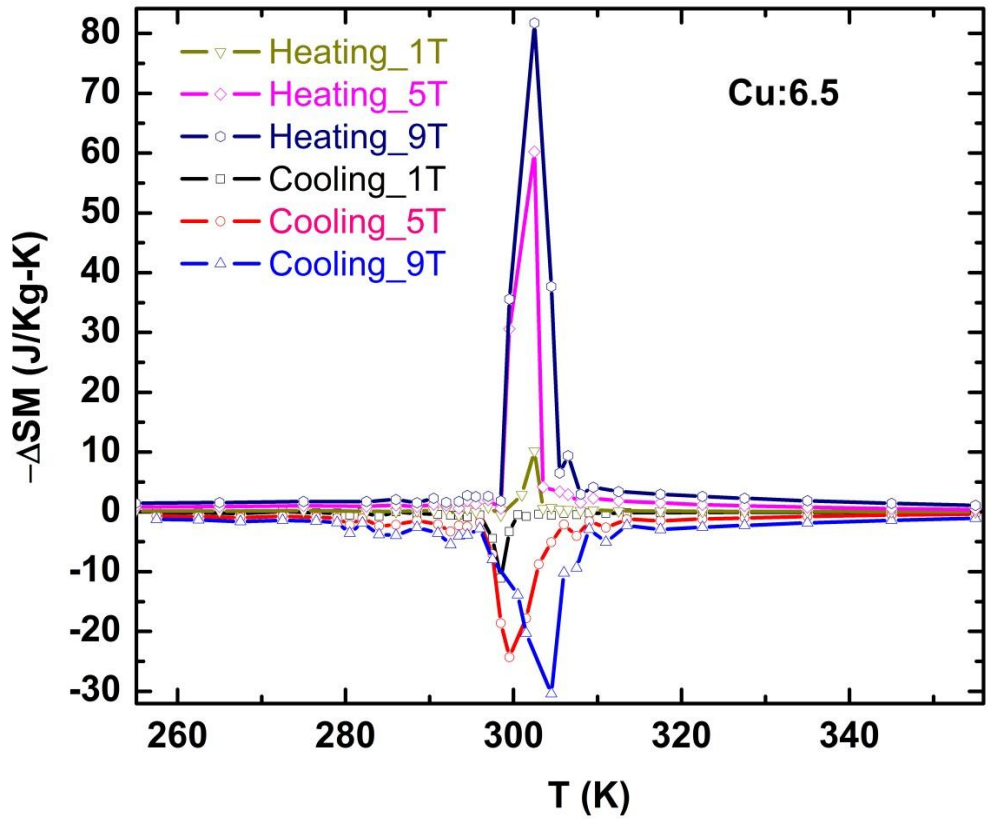

**Fig. 7.** Representative plot of magnetic entropy change with temperature for different magnetic field change, shown for Cu:6.5 alloy, for both heating as well as cooling cycle.

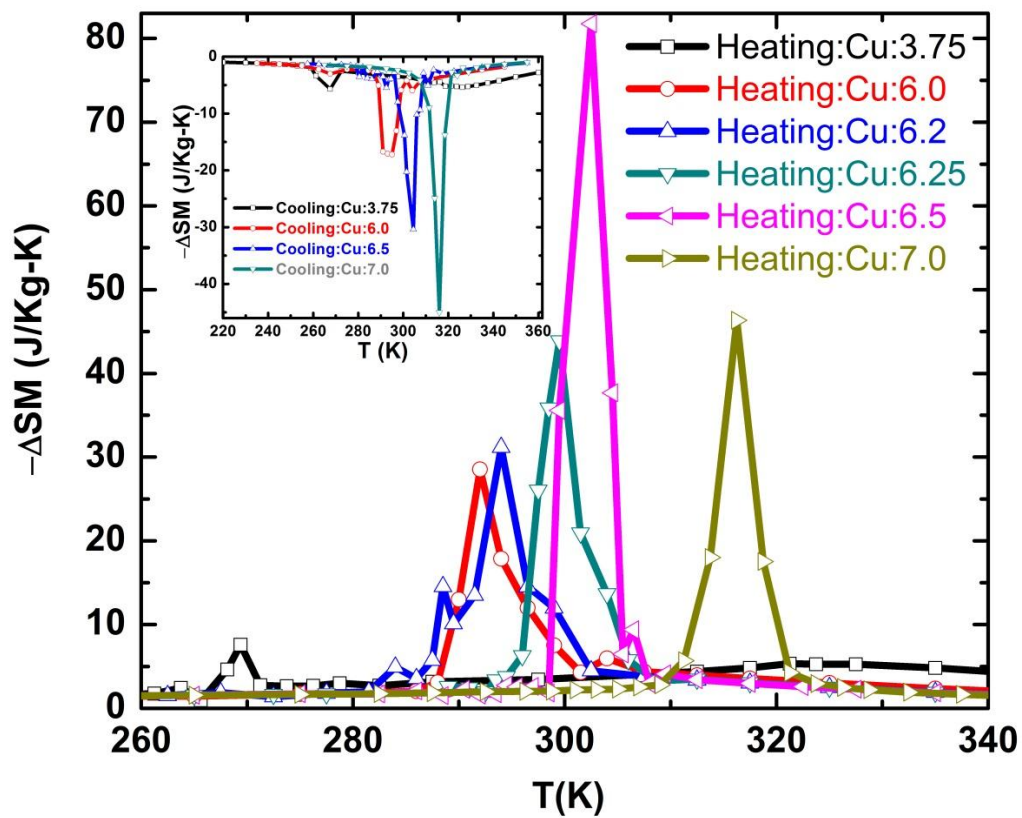

**Fig. 8.** Magnetic entropy calculated for ΔH = 9 T as function of Temperature (K) for different Cu concentrations for heating cycle and the inset shows the same for cooling cycle.

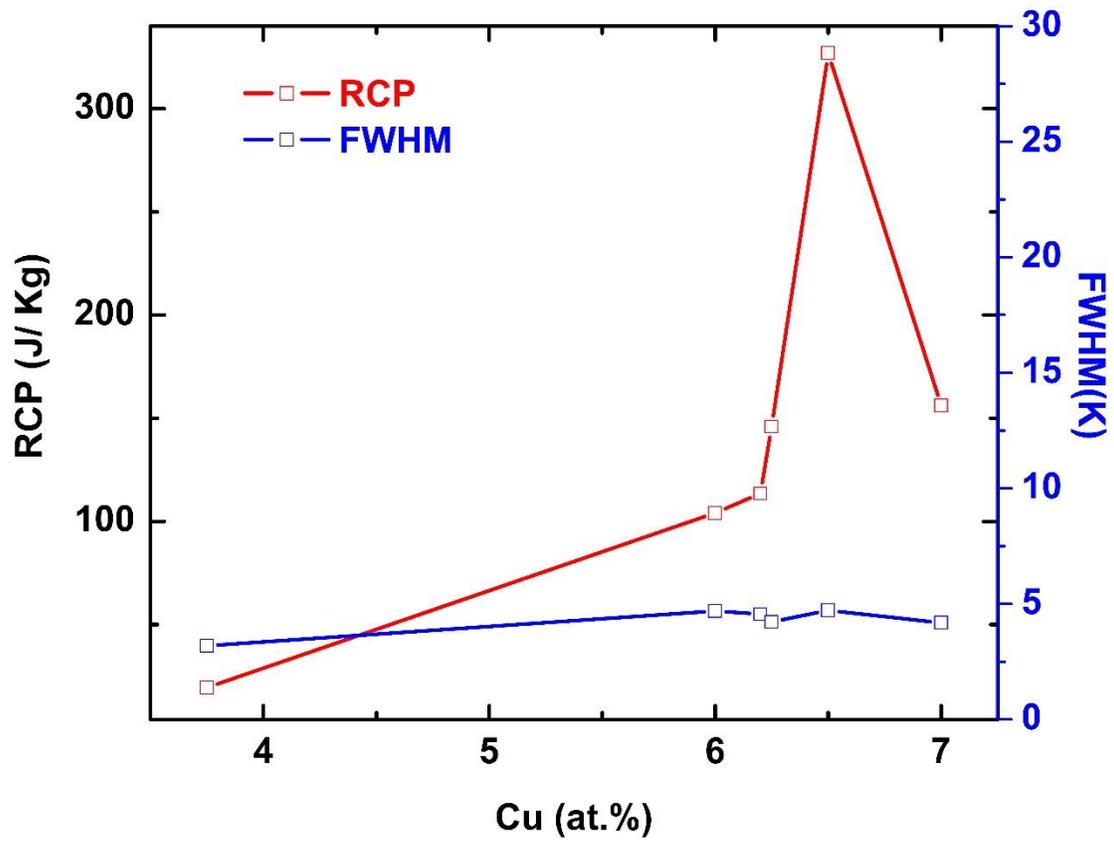

**Fig. 9.** Refrigerant capacity (RCP) and FWHM of $\Delta SM$ peak as a function of Cu concentration shown for magnetic field change of 9 T.

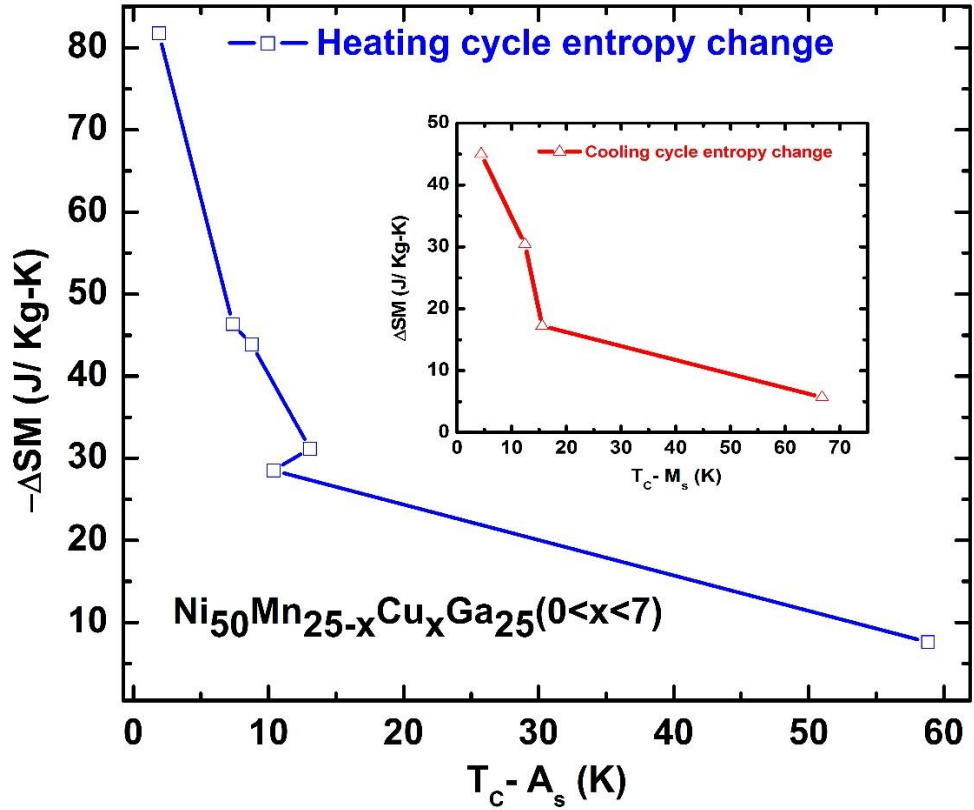

**Fig. 10.** Dependence of magnetic entropy change with ($T_C$ - $A_s$) for heating cycle and the inset shows the same for cooling cycle with ($T_C$ - $M_s$) for $Ni_{50}Mn_{25-x}Cu_xGa_{25}$ (0 < x < 7) alloys.